\documentclass{article}
\usepackage{spconf,amsmath,graphicx,hyperref}
\usepackage{amssymb,amsfonts}
\usepackage{algorithm}
\usepackage{algpseudocode}
\usepackage{booktabs} 
\usepackage{caption}  
\usepackage{subcaption}
\usepackage{multirow}
\usepackage{xcolor}
\usepackage{enumitem}

\title{ViTex: Visual Texture Control for Multi-Track Symbolic Music Generation via Discrete Diffusion Models}

\name{Xiaoyu Yi$^{\star\dagger}$ \qquad Qi He$^{\dagger}$  \qquad Gus Xia$^{\dagger}$ \qquad Ziyu Wang$^{\# \dagger }$}
  
\address{$^{\star}$ School of EECS, Peking University \\
      $^{\dagger}$Music X Lab, MBZUAI \\
      $^{\#}$ Courant Institute of Mathematical Sciences, New York University}

\makeatletter
\renewcommand{\section}{\@startsection{section}{1}{0pt}%
  {5pt} 
  {5pt} 
  {\centering\Large\bfseries}}

\renewcommand{\subsection}{\@startsection{subsection}{2}{0pt}%
  {4pt}
  {3pt} 
  {\normalfont\bfseries}}
\makeatother

\begin{document}
\ninept
\topmargin=0mm
\maketitle
\begin{abstract}
In automatic music generation, a central challenge is to design controls that enable meaningful human-machine interaction. Existing systems often rely on extrinsic inputs such as text prompts or metadata, which do not allow humans to directly shape the composition. While prior work has explored intrinsic controls such as chords or hierarchical structure, these approaches mainly address piano or vocal-accompaniment settings, leaving multitrack symbolic music largely underexplored. We identify instrumentation, the choice of instruments and their roles, as a natural dimension of control in multi-track composition, and propose ViTex, a visual representation of instrumental texture. In ViTex, color encodes instrument choice, spatial position represents pitch and time, and stroke properties capture local textures. Building on this representation, we develop a discrete diffusion model conditioned on ViTex and chord progressions to generate 8-measure multi-track symbolic music, enabling explicit texture-level control while maintaining strong unconditional generation quality. The demo page and code are avaliable at \url{https://vitex2025.github.io/}.

\end{abstract}

\section{Introduction}

In automatic music generation, a key challenge is to design \textit{controls} that enable meaningful human-machine interaction. Many existing systems rely on extrinsic inputs such as text prompts or metadata (e.g., composer identity, pitch range), which do not allow humans to directly participate in the creation process \cite{mmt, figaro}. In pursuit of more intrinsic forms of control, prior research \cite{polydis, polyffusion, whole_song} has explored chord progressions, accompaniment textures, and hierarchical abstractions. However, these approaches focused mainly on piano-only or vocal-accompaniment scenarios, leaving the more complex domain of multi-track symbolic music generation largely underexplored.

When music expands from piano writing to multi-track composition, a new dimension of control emerges: instrumentation, the choice of instruments and their roles, which brings new color to the music. This corresponds to the \textit{polyphonic texture} we focus on in this paper: deciding which instruments play, when and in which pitch ranges, and how their textures combine into a coherent whole. The main challenge is that texture control is inherently high-dimensional, making it difficult to represent in an intuitive and interpretable way. Existing approaches either simplify this control to coarse instrument labels \cite{mmt,figaro}, or encode texture in latent variables that are hard to interpret and manipulate \cite{QnA,accomontage}. Others treat instrumentation implicitly through track-conditioned generation or infilling \cite{GETMusic, amt}, without offering top-down control over polyphonic texture.

\begin{figure}[t] 
    \centering
    \includegraphics[width=0.45\textwidth]{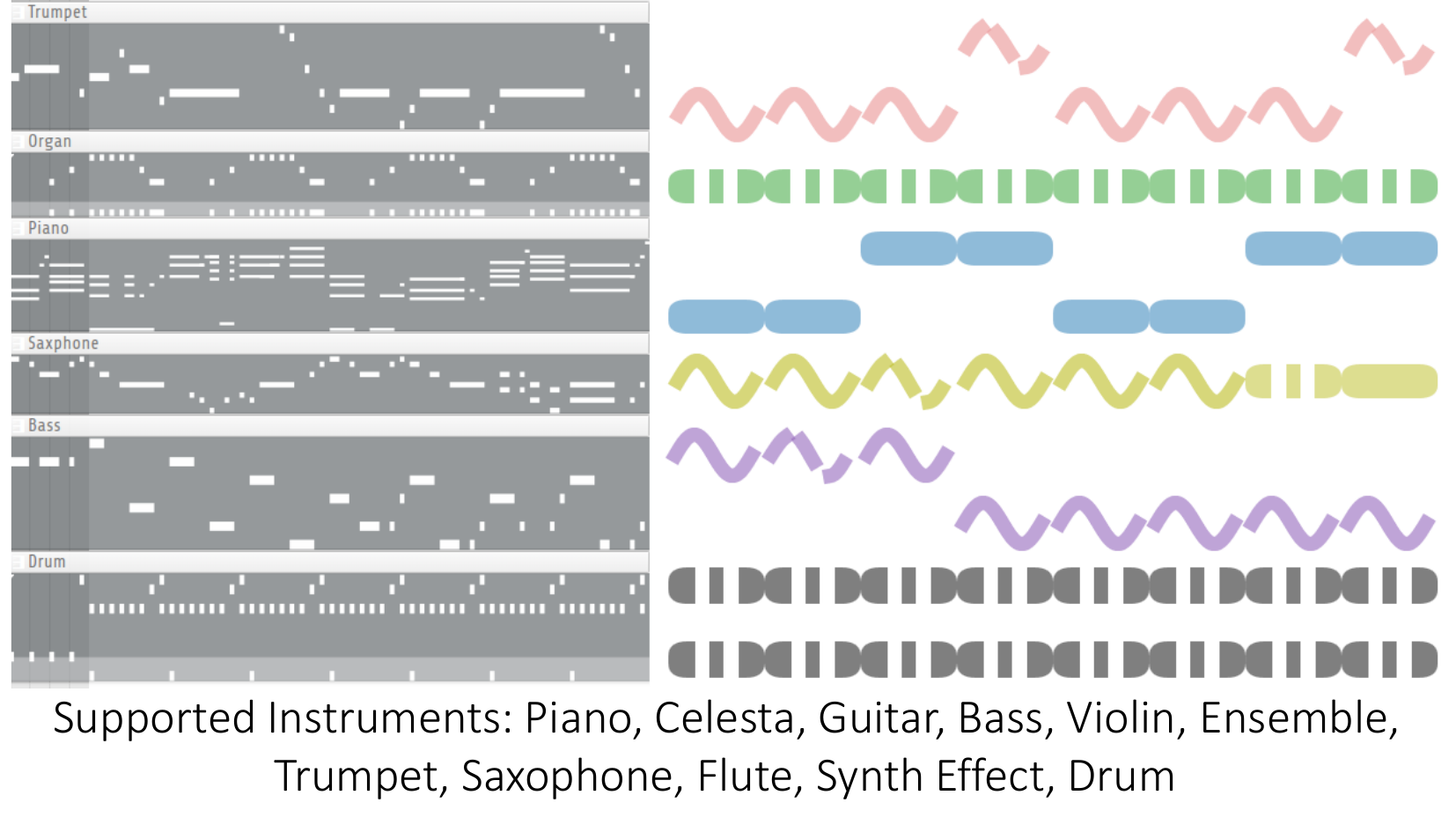} 
    \caption{Visualization of ViTex. Left: piano rolls of each track in a digital audio workstation (DAW). Right: the corresponding ViTex visualization. Different glyphs represent local textures. Horizontal axis indicates time; vertical positions within each track indicate the pitch.}
    \label{fig:vitex_illustration_new}
\end{figure}

To address this gap, we propose ViTex, a visual representation of instrumental texture designed for controllable multi-track music generation. An example is shown in Figure \ref{fig:vitex_illustration_new}. In ViTex, color encodes instrument choice, spatial position represents pitch and time, and stroke properties capture local textures. The current version supports multiple instruments, including drums, which are often omitted in texture modeling. This visual abstraction offers an intuitive yet expressive way to specify multi-track texture, a spirit it shares with Hyperscore \cite{hyperscore}, an earlier interface that enabled users to compose by drawing before the era of deep generative models and demonstrated the promise of such representations for intuitive human control. At the same time, our design is easily translatable into machine-friendly data suitable for conditioning modern generative models.

Building on ViTex, we develop a discrete diffusion model conditioned on the ViTex representation and chord progressions to generate 8-measure multi-track symbolic music. The model is trained to denoise a multi-track pianoroll from an absorbing state, with the ViTex representation and the chords incorporated as conditions. Both conditions are trained with classifier-free guidance \cite{cfg}, allowing the model to flexibly switch between conditional and unconditional generation. Experiments show that our model offers strong controllability, with generation quality on par with the baselines.

\section{Related Work}\nopagebreak

\subsection{Control in Symbolic Music Generation}
Control in symbolic music generation has addressed various aspects. In piano music, many works employ VAEs to extract latent representations that guide generation, such as EC$^2$-VAE \cite{EC2VAE} for rhythm, Polydis \cite{polydis} for chord and texture, and Music SketchNet \cite{Musicsketchnet} for pitch and rhythm. In the multitrack setting, however, control is typically simplified, such as by encoding instrument sets or global properties with tokens \cite{mmt,figaro,musecoco} or by providing partial music for inpainting \cite{amt,GETMusic}. Yet direct and interpretable track-level texture control remains unexplored, as existing latent-based approaches \cite{polydis} are difficult to interpret. Meanwhile, systems such as Draw and Listen \cite{Drawandlisten} and Hyperscore \cite{hyperscore} highlight the value of intuitive visualization and interaction by mapping sketches and curves to abstract musical ideas. Building on these insights, we propose ViTex, a representation that complements existing approaches by visualizing multitrack polyphonic texture while enabling intuitive control for generation tasks.

\subsection{Multitrack Symbolic Music Generation}
Recent multitrack symbolic music generation methods mostly adopt transformer architectures \cite{attention} to model note-level token sequences, either by directly modeling the sequence itself \cite{amt} or by jointly modeling note tokens with control tokens \cite{mmt,figaro,musecoco} that abstract simple musical conditions. GETMusic \cite{GETMusic}, in contrast, uses a diffusion-based approach, employing absorbing-state discrete diffusion \cite{D3PM} to generate from any source track to any target track. Some works focus on arrangement \cite{xiaoice,QnA,accomontage,popmag}, typically taking inputs with pitch information (e.g., melody or chords) and optionally incorporating information from other tracks to guide the arrangement. However, none of these approaches generate music conditioned on track-level polyphonic texture; at most, they allow control over instrument presence \cite{mmt,figaro,GETMusic,popmag} or simple statistical features of the instruments \cite{accomontage,QnA}. In practice, composers often prefer precise control, such as specifying that the bass plays a line in the lower register rather than merely being present. With our proposed ViTex capturing track-wise texture and combined with chord progression, our discrete diffusion model enables simultaneous control over both texture and harmony, addressing a gap in existing methods.

\subsection{Diffusion Techniques}
Diffusion models \cite{ddpm} have demonstrated remarkable capabilities across various generation tasks and have recently been extended to symbolic music generation \cite{diffusion_symbolic}. With classifier-free guidance \cite{cfg} enabling conditional control, Polyffusion \cite{polyffusion} allows manipulation of chords and texture, while whole-song generation \cite{whole_song} employs cascaded diffusion to produce complete pieces under both external and internal musical conditions. GETMusic \cite{GETMusic} further uses a discrete diffusion \cite{D3PM} model to handle multitrack music. These works highlight the potential of diffusion for modeling multitrack music and its strong controllability. Inspired by them, we adopt a absorbing state discrete diffusion framework \cite{D3PM} for multitrack symbolic music and leverage classifier-free guidance \cite{cfg} to incorporate our ViTex and chord-based controls.

\section{Method}
\label{sec:method}

\subsection{Data Representation}

\label{sec:data_representation}

We represent a multitrack symbolic music piece as a tensor, where each entry indicates the state of a given instrument at a specific timestep and pitch. Each state takes one of four values: silence, onset, sustain, or MASK. We use 16th-note resolution, considering 8-bar segments (32 beats) with 128 pitches across 11 instrument categories. Altogether, the piece can be denoted as a tensor $x \in \{0,1,2,3\}^{128 \times 128 \times 11}.$ Since drums are non-tonal, the drum track is represented by a separate single-channel pianoroll with shape $128 \times 128 \times 1$.

ViTex provides an intuitive representation of each track, derived from a rule-based method that also yields a machine-readable representation. For each track, we construct a low-resolution spatial feature map by aggregating non-overlapping 16×16 regions of the pianoroll, where each region corresponds to one beat along the temporal axis and 16 semitones along the pitch axis.

The texture of each region is determined via a rule-based analysis of the notes it contains. Inspired by \cite{couturier:tel-05046000}, we characterize musical texture along three perceptual dimensions: melodic, harmonic, and sustain. To distinguish melodic from harmonic texture, we count the number of simultaneously active notes within each region: regions containing simultaneous notes are labeled as harmonic and the rest as melodic. Sustain is determined by computing the non-silence ratio within each region, defined as the proportion of time during which at least one note is active relative to the region duration (one beat).
A region is labeled as sustain if this ratio is below 0.3, based on empirical observation.

Consequently, each track produces an $8 \times 8 \times 3$ texture feature map, and the 11 tracks are concatenated along the track axis, resulting in $y_{\text{ins}} \in \{0,1\}^{8 \times 8 \times 33}$, which encodes the track-wise instrumentation of the multitrack piece.

This design supports both intuitive visualization and a compact machine-readable representation. For visualization, each track is assigned a consistent color across time. Melodic regions are rendered as curves, while harmonic regions are rendered as rectangles. Non-sustained regions are indicated using dashed symbols to distinguish them from sustained texture. An example visualization is shown in Figure~\ref{fig:vitex_illustration_new}. The red trumpet track in measures 3-4 illustrates the contrast between non-sustained and sustained melodic texture. In contrast, the green organ and blue piano tracks highlight the corresponding distinction for harmonic texture. Beyond visualization, the resulting ViTex feature maps are also used as conditional inputs to guide the generative model.

We represent a chord by a $12 \times 3$ feature vector, with 12 pitch classes and 3 channels encoding root, chroma, and bass. Extracted at a per-beat resolution, this yields a chord representation $y_{\text{chd}} \in \{0,1\}^{32 \times 12 \times 3}$.

The drum track is treated differently from pitched instrument tracks. Each drum pitch is mapped to one of three frequency bands: low, mid, or high, using a fixed and empirically predefined mapping. For each band, we consider its rhythmic texture by characterizing whether drum hits occur, encoded with a binary indicator, and whether the activity is sustained over time. The latter directly reuses the sustain criterion defined above, which in the drum context corresponds to dense versus sparse rhythmic patterns. This results in an $8 \times 3 \times 2$ ViTex feature map for the drum track.

\subsection{Discrete Diffusion Model}
\label{sec:discrete_diffusion}

Following GETMusic \cite{GETMusic}, we adopt absorbing state discrete diffusion \cite{D3PM}. Given a categorical variable with $K$ possible states, we denote its value by $x$ and its one-hot representation by a column vector $v(x)$.  
The forward process is defined by a transition matrix $Q_t \in \mathbb{R}^{K \times K}$, where each entry  
$
[Q_t]_{mn} = q(x_t = m \mid x_{t-1} = n)
$ 
represents the transition probability from state $n$ to state $m$ at step $t$. We adopt absorbing-state diffusion, where at each timestep $t$ each entry either remains unchanged with probability $1-\beta_t$ or transits to MASK. $\beta_t$ is the corruption rate, which follows a cosine schedule.
 
\begin{figure}[t] 
    \centering
    \includegraphics[width=1.0\linewidth]{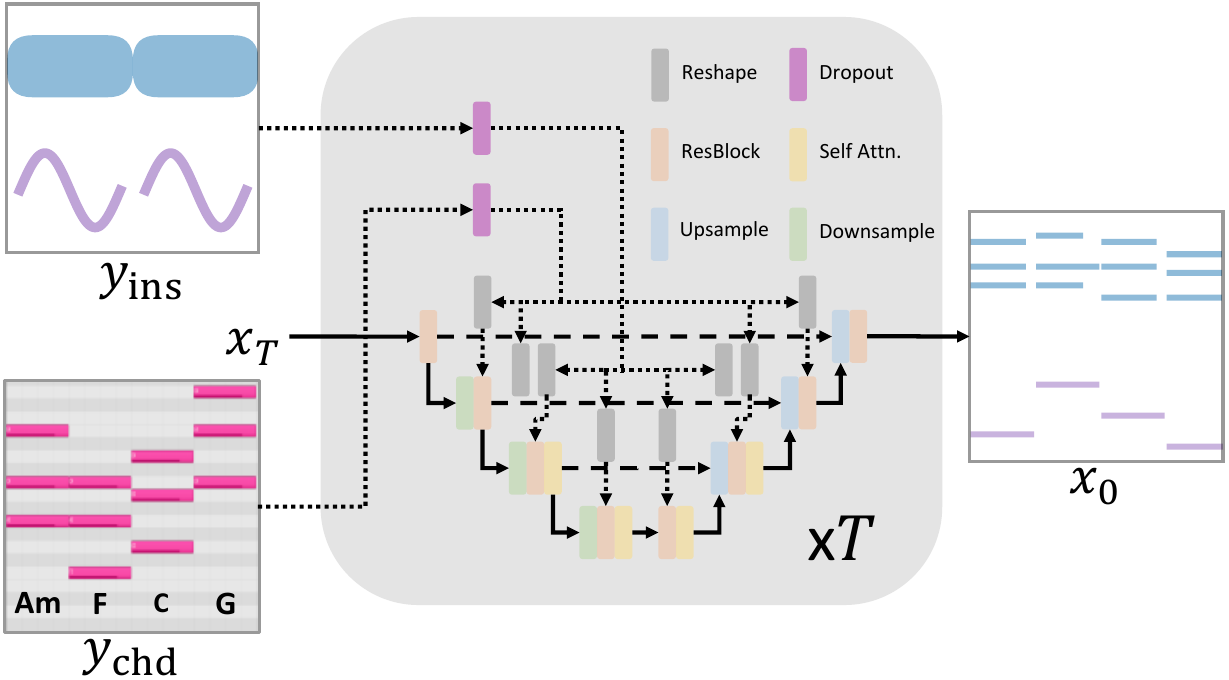} 
    \caption{Illustration of the UNet architecture with optional control from ViTex and chord progressions.}
    \label{fig:unet}
\end{figure}

The marginal and posterior distribution are
\begin{equation}
\label{eq:marginal}
  q(x_t \mid x_0) = v(x_t)^\top \hat{Q}_t \, v(x_0),\quad 
\text{where } \hat{Q}_t = \prod_{i=1}^t Q_i.  
\end{equation}
\begin{equation}
\label{eq:posterior}
\begin{aligned}
q(x_{t-1} \mid x_t, x_0) 
&= \frac{q(x_t \mid x_{t-1}, x_0) \, q(x_{t-1} \mid x_0)}{q(x_t \mid x_0)} \\[6pt]
&= \frac{\big(v(x_t)\!^\top\! Q_t\! \, v(x_{t-1})\!\big) \!
       \big(v(x_{t-1})\!^\top \! \hat{Q}_{t-1}\! \, v(x_0)\!\big)}
       {v(x_t)^\top \hat{Q}_t \, v(x_0)}.
\end{aligned}
\end{equation}
The reverse transition is parameterized as  
\begin{equation}
\label{eq:reverse}
  p_{\theta}(x_{t-1} \mid x_t) 
= \sum_{\hat{x}_0} q(x_{t-1} \mid x_t, \hat{x}_0) \, p_{\theta}(\hat{x}_0 \mid x_t).  
\end{equation}

We use a neural network $\epsilon_{\theta}(x_t, t)$ to predict the logits of the distribution $p_{\theta}(\hat{x}_0 \mid x_t)$ and directly optimize the $x_0$ prediction loss  
\begin{equation}
\label{eq:objective}
   \mathcal{L}_{\mathrm{pred}} = - \mathbb{E}_{t \sim \mathrm{U}(1,T), \, q(x_0)q(x_t \mid x_0)} \!\left[ \log p_\theta(x_0 \mid x_t) \right]. 
\end{equation}

\subsection{Guidance Mechanism}
We adopt classifier-free guidance \cite{cfg} to support conditional generation with ViTex and chord, as well as unconditional generation, which requires modifications to the above formulation. Our model takes two additional inputs, resulting in $\epsilon_\theta(x_t, t, y_1, y_2)$, where $y_1$ and $y_2$ correspond to $y_{\text{ins}}$ and $y_{\text{chd}}$, and are replaced with null with a probability of 0.5 during training. Accordingly, the training objective in Eq.~\ref{eq:objective} is adapted by replacing $p_\theta(x_0 \mid x_t)$ with $p_\theta(x_0 \mid x_t, y_1, y_2)$.

During inference, we interpolate between the conditional and unconditional predictions. Specifically, given instrumentation $y_{\text{ins}}$ and chord progression $y_{\text{chd}}$, with corresponding guidance strengths $\lambda_{\text{ins}}$ and $\lambda_{\text{chd}}$, the guidance is applied as\footnote{For brevity, we omit $x_t$ and $t$ in the notation.}
\begin{equation}
\begin{aligned}
\label{eq:guidance}
\hat{\epsilon}_\theta(y_{\text{ins}}, y_{\text{chd}}) &= 
\epsilon_\theta(\varnothing, \varnothing) \\ 
&\quad + \lambda_{\text{ins}} \Big(\epsilon_\theta(y_{\text{ins}}, \varnothing) - \epsilon_\theta(\varnothing, \varnothing)\Big) \\ 
&\quad + \lambda_{\text{chd}} \Big(\epsilon_\theta(\varnothing, y_{\text{chd}}) - \epsilon_\theta(\varnothing, \varnothing)\Big) \\
&\quad + \lambda_{\text{ins}} \lambda_{\text{chd}} \Big(\epsilon_\theta(y_{\text{ins}}, y_{\text{chd}}) 
- \epsilon_\theta(y_{\text{ins}}, \varnothing) \\
&\qquad - \epsilon_\theta(\varnothing, y_{\text{chd}}) + \epsilon_\theta(\varnothing, \varnothing) \Big).
\end{aligned}
\end{equation}

Given $x_t$, we first compute the logits of predicted distribution of $\hat{x}_0$ using the guidance formula in Eq.~\ref{eq:guidance} and sample an $\hat{x}_0$. 
If inpainting is applied \cite{repaint}, the sampled $\hat{x}_0$ is partially replaced by the known values in the masked regions. 
Finally, $x_{t-1}$ is sampled from the posterior in Eq.~\ref{eq:posterior}, completing a single reverse diffusion step described in Eq.~\ref{eq:reverse}.

\subsection{UNet Architecture}
\label{sec:unet}
To model $\epsilon_{\theta}(x_t, t, y_1, y_2)$, we adopt a standard UNet \cite{unet} with self-attention \cite{attention} to process $x_t$ and $t$. The conditional inputs are injected by element-wise addition of the feature maps derived from $y_1$ and $y_2$ to the corresponding latent feature maps in each UNet layer. Since the feature maps of $y_1$ and $y_2$ do not initially match the shape of the UNet latent feature maps, we first pass them through a reshape block, where the width dimension is aligned by direct stretching, while the height and channel dimensions are transformed using a small MLP to match the corresponding UNet feature map. After this alignment, the conditional feature maps can be added element-wise to the UNet latent feature maps at each layer. Our UNet has four levels, where $y_1$ is injected into the bottom two levels and $y_2$ into the middle two levels. Figure \ref{fig:unet} shows the architecture of the model.
\section{Experiment}
In this section, we compare ViTex with existing methods to highlight two key advantages: (1) it provides finer control over musical texture than existing control approaches, and (2) it achieves generation quality on par with state-of-the-art multi-track models.
\label{sec:exp}
\subsection{Dataset and Training Details}

Experiments are conducted on filtered subsets of the Lakh MIDI \cite{lakh} and Meta MIDI \cite{metamidi} datasets. We keep pieces in 4/4 meter, with tempos between 110 and 130 BPM, no key changes, at least 40 bars long, with at least five active tracks spanning more than three instrument categories, more than 50 notes, including drums and at least one of piano, guitar, or bass, while discarding files with excessive zero-duration notes. This results in 7,175 songs, split 90\%/10\% for training and testing at the song level. Data is processed by muspy \cite{muspy}. Models are trained using AdamW with a learning rate of 3e-4 and batch size 100, under a cosine decay schedule on 4 H100 GPUs.

\subsection{Baselines}

We use Q\&A \cite{QnA} as a baseline for conditional generation. Q\&A takes a polyphonic score as input, allowing us to provide chord sequences, and, for each track, a pitch distribution and note density encoded into a latent representation of track texture. To make our ViTex compatible with Q\&A and enable track-wise texture control, we design two conversion strategies: 
\begin{enumerate}[itemsep=0pt,topsep=1pt]
    \item Q\&A-1 (strict): Chords are used directly as a polyphonic score; track-wise pitch distributions and note densities are extracted from a mock multitrack piece generated conditioned on ViTex and the chord progression.
    \item Q\&A-2 (relaxed): Starting from Q\&A-1, random non-chord tones are added and the resulting pitch distributions and note densities are smoothed.
\end{enumerate}
These two variants serve as our conditional generation baselines. This choice and adaptation is reasonable because our model supports very fine-grained track-wise texture control, and Q\&A is the closest available baseline in terms of control granularity to the best of our knowledge.

We use the Anticipatory Music Transformer (AMT)\cite{amt} and the Multitrack Music Transformer (MMT)\cite{mmt} as baselines for unconditional generation and continuation. Moreover, MMT supports specifying which tracks should be present, making it suitable as a baseline for conditional generation in subjective listening experiments alongside Q\&A.
\subsection{Objective Evaluation}
We conduct conditional generation to evaluate controllability and quality. Using 1,000 randomly sampled test segments with their ViTex representations and chord progressions, we generate outputs with our model, Q\&A-1, and Q\&A-2 \cite{QnA}. For each sample, we compute the cosine similarity between the generated and ground-truth ViTex feature maps to evaluate Instrumentation Accuracy (IA), and between the generated and ground-truth chord progressions to evaluate Chord Accuracy (CA).
\begin{table}[t]
\centering
\caption{Objective evaluation for conditional generation.}
\label{tab:cond}
\setlength{\tabcolsep}{4pt}
\begin{tabular}{l c c c c c c}
\toprule
\multirow{2}{*}{Model} & \multicolumn{3}{c}{Instrumentation Control} & \multicolumn{2}{c}{Chord Control} & Quality \\
\cmidrule(lr){2-4} \cmidrule(lr){5-6} \cmidrule(l){7-7} 
 & IA$\uparrow$ & OAD$\uparrow$ & OAIOI$\uparrow$ & CA$\uparrow$ & OAP$\uparrow$ & DOA$\uparrow$ \\
\midrule
Q\&A-1\cite{QnA} & 0.584 & 0.135 & 0.451 & 0.607 & 0.450 & 0.188 \\
Q\&A-2\cite{QnA} & 0.299 & 0.082 & 0.110 & 0.043 & 0.253 & 0.111 \\
Ours &  \textbf{0.600} & \textbf{0.626} & \textbf{0.494} & \textbf{0.875} & \textbf{0.731} & \textbf{0.296} \\
\bottomrule
\end{tabular}
\end{table}
\begin{table}[t]
\centering
\caption{Objective evaluation for unconditional generation.\\ \centering $^*$Closer to ground truth is better. \\ For PCE and GPS, + and - is for deviation.}
\label{tab:uncond}
\fontsize{9pt}{11pt}\selectfont 
\begin{tabular}{lccc}
\toprule
Model & PCE$^*$ & GPS$^*$ & DOA $\uparrow$ \\
\midrule
Ground Truth & 1.741 & 0.804 & 0.303 \\
\midrule 
MMT\cite{mmt} & \textbf{+0.103} & +0.080 & 0.171 \\
AMT\cite{amt} & -0.317 &  +0.174 & 0.278 \\
Ours & -0.174 & \textbf{+0.050} & \textbf{0.307} \\
\bottomrule
\end{tabular}
\end{table}
Additionally, we evaluate the averaging Overlapped Area of distributions for note duration (OAD), inter-onset interval (OAIOI), and pitch (OAP) following previous works \cite{popmag}. Larger values of OAD, OAIOI, and OAP indicate greater overlap between the distributions of the corresponding features in the generated samples and the ground truth, demonstrating improved controllability. For generation quality, we adopt the Degree of Arrangement (DOA) metric proposed in \cite{accomontage}, where a higher DOA indicates a better-quality arrangement. As shown in Table \ref{tab:cond}, our model outperforms the baselines in both controllability and quality, demonstrating its ability to generate high-quality music while effectively adhering to the specified controls.

In addition to conditional generation, we conduct unconditional generation experiments to further evaluate the quality of generated samples. For each model, we generate 1,000 segments from scratch and compare them with 1,000 randomly sampled segments from the test set. We compute DOA, where higher values indicate better arrangement quality, and, following prior work \cite{mmt,jazz_transformer}, also calculate Pitch Class Entropy (PCE) to assess tonal clarity and Grooving Pattern Similarity (GPS) to measure rhythmicity, with values closer to the ground truth considered better. Absolute values are reported for DOA, while deviations from the ground truth are reported for PCE and GPS. As shown in Table \ref{tab:uncond}, our model outperforms the baselines on GPS and DOA, slightly underperforms on PCE, and overall demonstrates strong capability in generating high-quality music from scratch, comparable to existing methods.

\begin{figure}
    \centering
    \begin{subfigure}{0.8\linewidth}
        \centering
        \includegraphics[width=\linewidth]{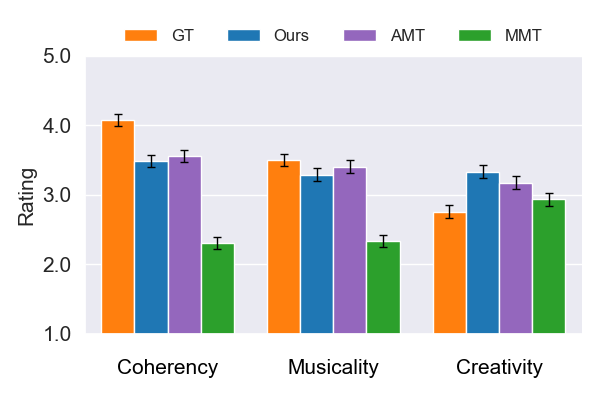}
        \caption{Results of music continuation given prompts.}
        \label{fig:subjeval1}
    \end{subfigure}
    \begin{subfigure}{0.8\linewidth}
        \centering
        \includegraphics[width=\linewidth]{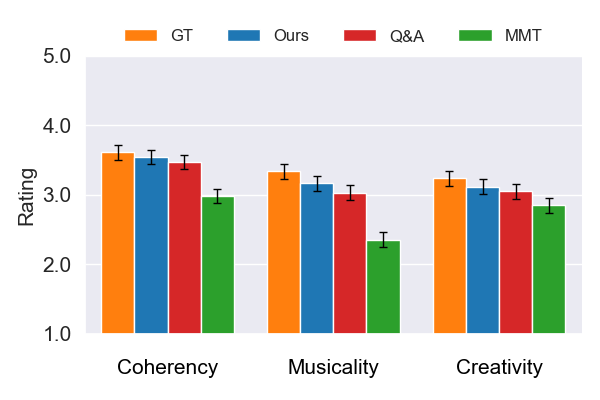}
        \caption{Results of music geneartion given instruments.}
        \label{fig:subjeval2}
    \end{subfigure}
    \caption{Subjective evaluation results with mean ratings and within-subject confidence intervals.}
    \label{fig:subjeval}
\end{figure}

\subsection{Subjective Evaluation}

We conducted a listening test with 31 participants to evaluate subjective quality. The study involved two tasks: music continuation given prompts and music generation given instruments. In both tasks, participants listened to outputs from our model, the baselines, and the ground truth, and rated each sample on a 5-point Likert scale for the following criteria: \textit{Coherency} on how well the continuation adheres to the style of the prompt, \textit{Musicality} on the overall quality and structural clarity of the music, and the sample's \textit{Creativity}. For the continuation task, MMT\cite{mmt} and AMT\cite{amt} served as baselines; six test-set compositions were randomly selected, and their first two bars were used as prompts for each model to generate continuations. For the generation given instruments task, MMT\cite{mmt} and Q\&A\cite{QnA} were baselines; four test-set compositions were randomly selected, and ViTex representations along with chord progressions were provided to our model and Q\&A, while MMT received only instrument labels. 

Figure \ref{fig:subjeval} shows the results, where bar height represents mean ratings and error bars indicate confidence intervals computed using a two-way ANOVA without replication. In the music continuation task, our model achieves slightly lower Coherency and Musicality scores than AMT, but higher Creativity, and all metrics surpass MMT. This is reasonable, as AMT is specifically trained for infilling tasks. That our model performs comparably to AMT demonstrates that, under ViTex guidance, it can effectively follow the prompt, indicating that ViTex successfully captures instrumentation information. In the music generation task, our model outperforms all baselines across the evaluated metrics, highlighting both the quality and controllability of the generated music.

\section{Conclusion}
We propose ViTex, a representation that visualizes track-wise texture and enables novel fine-grained control over instrumentation. By combining ViTex with chord progression conditioning within a discrete diffusion framework, our approach supports both conditional and unconditional multitrack symbolic music generation with precise texture control. Objective and subjective evaluations demonstrate that our model provides strong controllability while maintaining generation quality on par with state-of-the-art baselines.

\bibliographystyle{IEEEbib}
\bibliography{strings,refs}

\end{document}